\documentclass[amsmath,amssymb,twocolumn,showpacs,superscriptaddress,groupedaddress,aps,prl,10pt,nofootinbib]{revtex4-1}  
\usepackage{graphicx}  
\usepackage{dcolumn}   
\usepackage{bm}        
\usepackage{bbm}
\usepackage{cleveref}
\usepackage{ulem}

\begin{document}

\title{Photon-photon scattering at the high-intensity frontier}

\author{Holger Gies} 
\email{holger.gies@uni-jena.de}
\affiliation{Theoretisch-Physikalisches Institut, Abbe Center of Photonics,
Friedrich-Schiller-Universit\"at Jena, Max-Wien-Platz 1, D-07743 Jena, Germany \\ 
Helmholtz-Institut Jena, Fr\"obelstieg 3, D-07743 Jena, Germany}
\author{Felix Karbstein} 
\email{felix.karbstein@uni-jena.de}
\affiliation{Theoretisch-Physikalisches Institut, Abbe Center of Photonics,
Friedrich-Schiller-Universit\"at Jena, Max-Wien-Platz 1, D-07743 Jena, Germany \\ 
Helmholtz-Institut Jena, Fr\"obelstieg 3, D-07743 Jena, Germany}
\author{Christian Kohlf\"urst} 
\email{christian.kohlfuerst@uni-jena.de}
\affiliation{Theoretisch-Physikalisches Institut, Abbe Center of Photonics,
Friedrich-Schiller-Universit\"at Jena, Max-Wien-Platz 1, D-07743 Jena, Germany \\ 
Helmholtz-Institut Jena, Fr\"obelstieg 3, D-07743 Jena, Germany}
\author{Nico Seegert} 
\email{nico.seegert@gmx.net}
\affiliation{Theoretisch-Physikalisches Institut, Abbe Center of Photonics,
Friedrich-Schiller-Universit\"at Jena, Max-Wien-Platz 1, D-07743 Jena, Germany \\ 
Helmholtz-Institut Jena, Fr\"obelstieg 3, D-07743 Jena, Germany}
                             
\date{\today}

\begin{abstract}
The tremendous progress in high-intensity laser technology and the establishment of dedicated high-field laboratories in recent years have paved the way towards a first observation of quantum vacuum nonlinearities at the high-intensity frontier.
We advocate a particularly prospective scenario, where three synchronized high-intensity laser pulses are brought into collision, giving rise to signal photons, whose frequency and propagation direction differ from the driving laser pulses, thus providing various means to achieve an excellent signal to background separation.
Based on the theoretical concept of vacuum emission, we employ an efficient numerical algorithm which allows us to model the collision of focused high-intensity laser pulses in unprecedented detail. We provide accurate predictions for the numbers of signal photons accessible in experiment.
Our study is the first to predict the precise angular spread of the signal photons, and paves the way for a first verification of quantum vacuum nonlinearity in a well-controlled laboratory experiment at one of the many high-intensity laser facilities currently coming online.
\end{abstract}

\maketitle

\paragraph{Introduction}

The quantum vacuum is characterized by the omnipresence of fluctuations of the underlying theory's particle degrees of freedom -- in quantum electrodynamics (QED): electrons/positrons and photons.
As electromagnetic fields couple to charges, the fluctuations of virtual charged particles can mediate effective interactions among electromagnetic fields \cite{Euler:1935zz,Heisenberg:1935qt,Weisskopf}, thereby invalidating one of the cornerstones of Maxwell's classical theory of electrodynamics, namely the celebrated superposition principle for electromagnetic fields in vacuum.
However, having no classical analogue, such {\it vacuum nonlinearities} are typically rather elusive in experiment; cf. the extremely small cross section for direct light-by-light scattering \cite{Euler:1935zz,Karplus:1950zz} mediated by an electron-positron loop. Nevertheless, these early studies of photon scattering, most notably the Heisenberg-Euler effective action \cite{Heisenberg:1935qt,Weisskopf,Schwinger:1951nm}, have formed a backbone of the evolution of modern quantum field theory.

The worldwide efforts in establishing dedicated laser facilities at the high-intensity frontier coming online just now, such as 
CILEX 
 \cite{CILEX}, 
CoReLS 
\cite{CoReLS}, 
ELI 
\cite{ELI} and
SG-II \cite{SG-II}
suggest a particularly promising route towards the first verification of QED vacuum nonlinearities in a laboratory experiment with macroscopically controllable electromagnetic fields.
These activities have stimulated numerous theoretical proposals, where the strong macroscopic electromagnetic fields of focused high-intensity laser pulses are employed to trigger interaction processes which have no analogue in classical electrodynamics; see the reviews \cite{Dittrich:2000zu,Marklund:2008gj,Dunne:2008kc,Heinzl:2008an,DiPiazza:2011tq,Battesti:2012hf,King:2015tba,Karbstein:2016hlj,Inada:2017lop} and references therein.
Aiming at performing such a discovery experiment with state-of-the-art technology, {\it all-optical signatures} of vacuum nonlinearity seem most promising.
This class of signatures encompasses fluctuation mediated interaction processes, where both the microscopic origin of the electromagnetic fields driving the effect and the signal itself are photons.

A prominent optical signature of QED vacuum nonlinearity is vacuum birefringence \cite{Toll:1952,Baier,BialynickaBirula:1970vy,Adler:1971wn}, predicted to be experienced by probe photons traversing a strong-field region. 
While already actively being searched for in experiments using macroscopic magnetic fields in combination with continuous-wave lasers and high-finesse cavities \cite{Cantatore:2008zz,Berceau:2011zz,Fan:2017fnd}, various recent theoretical studies have emphasized the possibility of its verification in an all-optical experiment, colliding an X-ray \cite{Heinzl:2006xc,DiPiazza:2006pr,Dinu:2013gaa,Karbstein:2015xra,Schlenvoigt:2016,Karbstein:2016lby} or gamma-ray probe \cite{Kotkin:1996nf,Nakamiya:2015pde,Ilderton:2016khs,King:2016jnl,Bragin:2017yau} with a high-intensity laser pulse.
Other theoretical proposals have focused, e.g., on vacuum nonlinearity induced photon scattering phenomena in laser pulse collisions \cite{Moulin:2002ya,Lundstrom:2005za,Monden:2011,King:2012aw,wir}, interference effects \cite{Hatsagortsyan:2011,King:2013am,Tommasini:2010fb}, laser mode self-mixing \cite{Paredes:2014oxa}, quantum reflection \cite{Gies:2013yxa}, higher-harmonic generation in an electromagnetized vacuum \cite{BialynickaBirula:1981,Kaplan:2000aa,Valluri:2003sp,Fedotov:2006ii,Marklund:2006my,DiPiazza:2005jc,King:2014vha,Bohl:2015uba}, photon splitting \cite{BialynickaBirula:1970vy,Adler:1970gg,Adler:1971wn,Lee:1998hu,Papanyan:1971cv,Stoneham:1979,Baier:1986cv,Adler:1996cja,DiPiazza:2007yx} and photon merging \cite{Yakovlev:1966,DiPiazza:2007cu,Gies:2014jia,Gies:2016czm}.

Our work builds on the recent observation that optical signatures
of quantum vacuum nonlinearities in inhomogeneous electromagnetic
fields can be efficiently analyzed by reformulating them as {\it
  vacuum emission} processes \cite{Karbstein:2014fva}.  In this
picture, signatures of vacuum nonlinearities are encoded in {\it
  signal photons} induced in the interaction volume of the
high-intensity laser pulses driving the effect.  The latter are
formally described as classical background fields, which is
well-justified, because laser beams propagating in vacuum are
optimal examples for coherent macroscopic fields.

Upon combination with an efficient numerical algorithm, our approach
facilitates quantitative theoretical studies for a wide variety of
experimentally realistic field configurations at unprecedented
  accuracy \cite{wir}.  In this letter, we investigate photon-photon
scattering in the collision of three synchronized high-intensity laser
pulses modeled as pulsed paraxial Gaussian beams. For an experimental upper bound, cf. Ref.~\cite{Bernard:2010dx}. More specifically,
we focus on two different experimental scenarios, both of which will
become possible in the near future at various high-intensity
facilities. Our approach is based on a locally-constant-field approximation of the Heisenberg-Euler effective
  Lagrangian.  In contrast to previous studies, our formalism does not
  require any additional {\it ad hoc} approximations: upon
  specification of the macroscopic electromagnetic fields of the
  high-intensity laser beams, the signal photons and their
  polarization and propagation properties are unambiguously
  predicted.

\paragraph{Formalism}

Far outside the  interaction volume, the differential number of signal photons of polarization $p$ arising from the effective interaction of macroscopic electromagnetic fields in the quantum vacuum can be compactly represented as \cite{Karbstein:2014fva}
\begin{equation}
{\rm d}^3N_{(p)}(\vec{k})=\frac{{\rm d}^3k}{(2\pi)^3}\bigl|{\cal S}_{(p)}(\vec{k})\bigr|^2 \,, \label{eq:d3Np}
\end{equation}
where ${\cal S}_{(p)}(\vec{k})$ is the zero-to-single signal photon transition amplitude,
induced by the effective coupling of the laser fields via vacuum fluctuations. For a polarization-insensitive measurement, the polarization-summed differential number of signal photons is ${\rm d}^3N(\vec{k})=\sum_p{\rm d}^3N_{(p)}(\vec{k})$.

For locally constant fields, i.e., fields varying on scales much larger than the Compton wavelength of the electron $\lambdabar_{\rm C}\approx3.86 \cdot 10^{-13}\,{\rm m}$, it suffices to describe the nonlinear interactions of the strong fields in terms of the Heisenberg-Euler effective Lagrangian $\mathcal{L}_{\text{HE}}$ \cite{Heisenberg:1935qt,Weisskopf,Schwinger:1951nm}. Then, we obtain
\begin{equation}
{\cal S}_{(p)}(\vec{k})
 =\frac{\epsilon^{*\nu}_{(p)}(\vec{k})}{\sqrt{2k^0}} \, 2{\rm i} k^\mu\int{\rm d}^4 x\,{\rm e}^{{\rm i}kx}\,\frac{\partial{\cal L}_\text{HE}}{\partial F^{\mu\nu}}\bigg|_{F\to F(x)}\,, \label{eq:Sp}
\end{equation}
with $k^0\equiv|\vec{k}|$, using the conventions detailed in Ref.~\cite{wir}.

In spherical coordinates $\vec{k}={\rm k}\hat{\vec{e}}_k$, the propagation directions of the signal photons can be expressed as 
$\hat{\vec{e}}_k=(\cos\varphi\sin\vartheta,\sin\varphi\sin\vartheta,
\cos\vartheta)$, and the unit vectors perpendicular to $\hat{\vec{e}}_k$ as
$\hat{\vec{e}}_{\beta}=\sin\beta\,\hat{\vec{e}}_k|_{\vartheta=\frac{\pi}{2},\varphi\to\varphi+\frac{\pi}{2}}+\cos\beta\,\hat{\vec{e}}_k|_{\vartheta\to\vartheta+\frac{\pi}{2}}$, with $\beta$ parameterizing all possible orientations.
Hence, $\epsilon^\mu_{(p)}(\vec{k}):=(0,\hat{\vec{e}}_{\beta=\beta_0+\frac{\pi}{2}(p-1)})$ with $p\in\{1,2\}$ span the linear polarizations of signal-photons propagating along $\hat{\vec{e}}_k$; $\beta_0$ fixes the polarization basis.

Using these definitions, at one loop and leading order in $\frac{eF}{m_e^2}\ll1$, Eq.~\eqref{eq:Sp} becomes \cite{wir}
\begin{align}
&{\cal S}_{(p)}(\vec{k})
 = \frac{1}{\rm i}\frac{e}{4\pi^2}\frac{m_e^2}{45}\sqrt{\frac{\rm k}{2}}\Bigl(\frac{e}{m_e^2}\Bigr)^3 \int{\rm d}^4 x\, {\rm e}^{{\rm i}{\rm k}(\hat{\vec{e}}_k\cdot\vec{x}-t)} \nonumber\\
&\,\quad\times\Bigl\{4\bigl[\hat{\vec{e}}_{\beta_0+\frac{\pi}{2}(p-1)}\cdot\vec{E}(x)-\hat{\vec{e}}_{\beta_0+\frac{\pi}{2}p}\cdot\vec{B}(x)\bigr] {\cal F}(x) \nonumber\\
&\,\quad\ \ +7\bigl[\hat{\vec{e}}_{\beta_0+\frac{\pi}{2}(p-1)}\cdot\vec{B}(x)+\hat{\vec{e}}_{\beta_0+\frac{\pi}{2}p}\cdot\vec{E}(x)\bigr] {\cal G}(x)\Bigr\} \,,
\label{eq:Spk}
\end{align}
where ${\cal F}(x)=\frac{1}{2}[\vec{B}^2(x)-\vec{E}^2(x)]$ and ${\cal
  G}(x)=-\vec{B}(x)\cdot\vec{E}(x)$.  We emphasize that
  Eq.~\eqref{eq:Spk} is very generic, and allows for the polarization
  sensitive study of signal photon emission from essentially all macroscopic
  electromagnetic field configurations available in the laboratory. Its evaluation is conceptually
  remarkably simple and calculationally straightforward.  In comparison,
  established approaches are typically based on a direct solution of the
  nonlinear wave equation, thereby necessitating the numerical
  solution of partial differential equations \cite{King:2012aw,King:2014vha,Bohl:2015uba,Domenech:2016xx,Carneiro:2016qus}. By contrast, our approach
  requires only an accurate implementation of fast Fourier transforms, and is thus easy-to-use. 
  The advantages of our formalism are
  particularly pronounced when aiming at quantitatively precise results
  in full 3+1 dimensional spacetime.

For a study of QED vacuum nonlinearity with high-intensity laser beams, the electric and magnetic fields entering Eq.~\eqref{eq:Spk} are given by
$\vec{B}(x)=\sum_b \vec{B}_b(x)$ and $\vec{E}(x)=\sum_b \vec{E}_b(x)$, where the sums are over the number of laser beams $b$ driving the effect.
In this letter, we consider a three-beam scenario with $b\in\{1,2,3\}$.

\paragraph{Beam modeling}

As a decisive new step, we compute the scattering for a realistic beam
description. While our method can be applied to any form of the beams,
we consider -- for definiteness -- the high-intensity laser fields to
be well-described as linearly polarized paraxial Gaussian beams
supplemented by a finite Gaussian pulse duration
\cite{Siegman,Karbstein:2015cpa}.  A given laser beam is thus fully
characterized by its propagation direction $\hat{\vec{e}}_{\kappa_b}$,
its field amplitude profile ${\cal E}_b(x)$, and its polarization unit
vector $\hat{\vec{e}}_{E_b}$, fulfilling
$\hat{\vec{e}}_{E_b}\cdot\hat{\vec{e}}_{\kappa_b}=0$.  At leading paraxial
order, the electric and magnetic fields
of each beam are given by $\vec{E}_b(x)={\cal
  E}_b(x)\hat{\vec{e}}_{E_b}$ and $\vec{B}_b(x)={\cal
  E}_b(x)\hat{\vec{e}}_{B_b}$, with
$\hat{\vec{e}}_{B_b}=\hat{\vec{e}}_{\kappa_b}\times\hat{\vec{e}}_{E_b}$.
Assuming each beam to be focused at $\vec{x}=\vec{x}_{0,b}$ and the temporal pulse amplitude profile to reach its maximum for $t=t_{0,b}$,
we define relative coordinates ${\rm t}_b:=t-t_{0,b}$, ${\rm z}_b:=\hat{\vec e}_{\kappa_b}\cdot\left(\vec{x}-\vec{x}_{0,b}\right)$ and $r_b:=\sqrt{(\vec{x}-\vec{x}_{0,b})^2-{\rm z}_b^2}$. While ${\rm z}_b$ is a longitudinal coordinate along the beam axis, $r_b$ corresponds to a radial coordinate relative to the beam axis, such that each field profile reads
\begin{equation}
 {\cal E}_b(x) = {\cal E}_{0,b}\,{\rm e}^{-\frac{({\rm z}_b-{\rm t}_b)^2}{(\tau_b/2)^2}}\, \frac{w_{0,b}}{w_b({\rm z}_b)}\, {\rm e}^{-\frac{r_b^2}{w_b^2({\rm z}_b)}}\, \cos\bigl(\Phi_b(x)\bigr) \,,
 \label{eq:E(x)}
\end{equation}
with phase
\begin{equation}
 \Phi_b(x) = \omega_b({\rm z}_b-{\rm t}_b)+\tfrac{{\rm z}_b}{{\rm z}_{R,b}}\tfrac{r_b^2}{w_b^2({\rm z}_b)}-\arctan\bigl(\tfrac{{\rm z}_b}{{\rm z}_{R,b}}\bigr)+\varphi_{0,b} \,.
 \label{eq:Phi(x)}
\end{equation}
Here, ${\cal E}_{0,b}$ is the peak field strength, $\omega_b=\frac{2\pi}{\lambda_b}$ the laser frequency and $\tau_b$ the pulse duration.
The transverse widening of the beam with ${\rm z}_b$ is encoded in the function $w_b({\rm z}_b) = w_{0,b} \sqrt{1+({\rm z}_b/{\rm z}_{{\rm R},b})^2}$, where $w_{0,b}$ denotes the beam waist, and ${\rm z}_{{\rm R},b}=\pi w_{0,b}^2/\lambda_b$ its Rayleigh range.
The second term in Eq.~\eqref{eq:Phi(x)} accounts for the curvature of the wavefronts as a function of ${\rm z}_b$ and $r_b$, 
$\arctan({\rm z}_b/{\rm z}_{R,b})$ is the Gouy phase shift, and $\varphi_{0,b}$ a constant phase.
The radial divergence of a given beam is $\theta_b\simeq w_{0,b}/{\rm z}_{{\rm R},b}$.

The peak field strength ${\cal E}_{0,b}$ is fully determined by the laser energy per pulse $W_b$, the pulse duration $\tau_b$ and the focus cross section $\sim\pi w_{0,b}^2$ \cite{Karbstein:2017jgh}:
\begin{equation}
 {\cal E}_{0,b}^2\approx8\sqrt{\frac{2}{\pi}}\frac{W_b}{\pi w_{0,b}^2\tau_b} \,. \label{eq:E0bquad}
\end{equation}

The minimum value of $w_0$ attainable in experiment depends on the focal length and the diameter of the focusing aperture, whose ratio defines the $f$-number $f^\#$, $w_{0,b}=f_b^\#\lambda_b$;
$f$-numbers as low as $f^\#=1$ can be realized experimentally \cite{Siegman}.
This we assume for all three laser beams, $w_{0,b}\equiv\lambda_b$, resulting in a radial beam divergence of $\theta_b\simeq1/\pi\approx18.24^\circ$.
A given laser beam is therefore fully parameterized by its propagation and polarization directions and the set $\{W_b,\tau_b,\omega_b\}$.

\paragraph{Results}

In general, the Fourier integral in Eq.~\eqref{eq:Spk} cannot be performed analytically. Hence, we employ the novel numerical algorithm devised in Ref.~\cite{wir} to carry out the Fourier integral, and upon taking the modulus squared of ${\cal S}_{(p)}(\vec{k})$, to evaluate the integrations over the signal-photon energies $\rm k$, and solid angle elements under consideration.

The number density of signal photons emitted in the direction $(\varphi,\vartheta)$
follows straightforwardly from Eq.~\eqref{eq:d3Np} by integrating over the signal photon energy $\rm k$, and reads
\begin{equation}
 \rho(\varphi,\vartheta):=\frac{1}{(2\pi)^3}\sum_{p=1}^2\int_0^\infty{\rm dk}\,\bigl|{\rm k} \, {\cal S}_{(p)}(\vec{k})\bigr|^2 \,. \label{eq:rhoomega}
\end{equation}
Note, that $\rho(\varphi,\vartheta)$ is independent of the
polarization base $\beta_0$.  The corresponding total number of signal
photons emitted into the full solid angle interval is given by
$N=\int_0^{2\pi}{\rm d}\varphi \int_{-1}^1{\rm
  d}\cos\vartheta\,\rho(\varphi,\vartheta)$.  For the scenarios
considered in this letter, the energy spectrum of the signal photons
can reliably be inferred from plane-wave predictions (cf. the detailed
discussion below), exhibiting a structure dictated by energy
conservation. The extraction and analysis of spectral and polarization
informations is, however, straightforward in our approach; cf., e.g.,
Ref.~\cite{Karbstein:2015xra} for an example.

For realistic estimates of signal photons $N$, we consider a high-field laboratory operating two equal high-intensity lasers, such as ELI-NP \cite{ELI}, which -- for a conservative estimate -- we assume to be of the one petawatt ($1\,{\rm PW}$) class, delivering pulses of energy $W=25\,{\rm J}$ and duration $\tau=25\,{\rm fs}$ at a frequency of $\omega=1.55\,{\rm eV}$ (wavelength $\lambda=800\,{\rm nm}$). As the laser energies enter our calculations only in terms of an overall factor, cf. Eq.~\eqref{eq:E0bquad},
our results can be straightforwardly rescaled to other laser energies.
For a three-beam experiment, one of the two laser beams is split into two, which is possible without significant loss, implying that $\{W,\tau,\omega\}\to\{W/2,\tau,\omega\}+\{W/2,\tau,\omega\}$.
As an additional handle for an efficient signal-to-background separation, we suggest frequency doubling to induce signal photons of distinct frequencies not present in the spectra of the high-intensity laser pulses. 
We conservatively estimate the energy loss for the conversion process preserving the pulse duration as ${\cal P}=50\%$ \cite{Marcinkevicius:2004}, such that $\{W,\tau,\omega\}\to\{W/2,\tau,2\omega\}$.
In general, the signal photon number $N$ scales as $(1-{\cal P})^{n}$, where $n$ is the number of frequency-doubled beams employed.
For concreteness, we consider two different collision geometries, both
of which were proposed by Ref.~\cite{Lundstrom:2005za} within a
plane-wave approach.  From here on, we assume perfect synchronization
and overlap, $t_{0,b}=0$ and $\vec{x}_{0,b}=0$, generalizations to
finite $t_{0,b}$ and $\vec{x}_{0,b}$ are straightforward within our
approach, cf. \cite{Karbstein:2016lby,wir}. {For instance, spatial
  displacements on the order of the beam waist size 
  have been found to deplete the signal photon numbers from two pulse collisions by a quantifiable $\mathcal{O}(1)$ factor \cite{wir}.}

\begin{figure}[ht]
\begin{center}
 \includegraphics[width=0.38\textwidth]{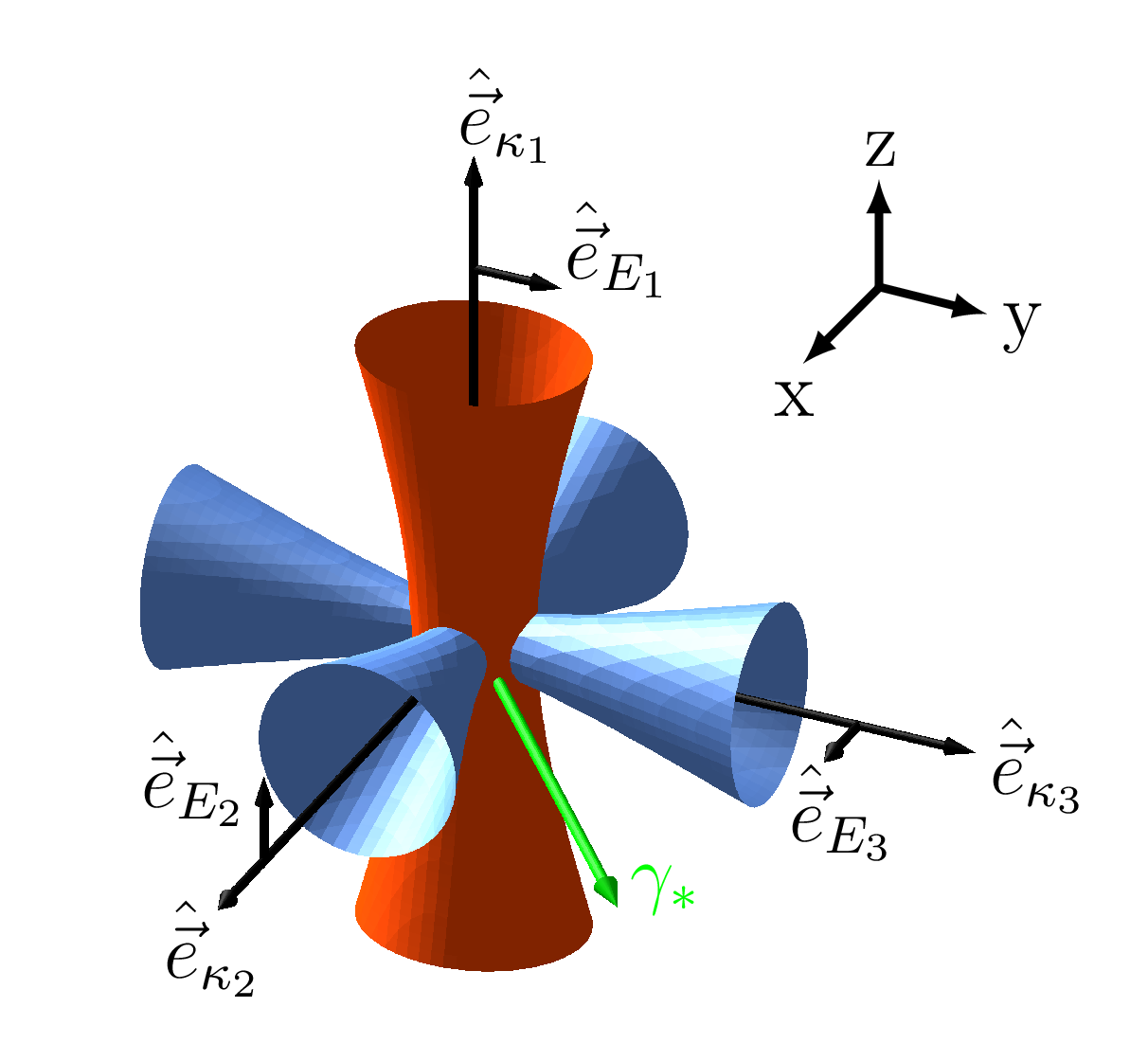}
\end{center} 
\vspace*{-0.5cm}
 \caption{Illustration of a three dimensional collision geometry, displaying the propagation directions $\hat{\vec{e}}_{\kappa_b}$ and polarization vectors $\hat{\vec{e}}_{E_b}$ of the high-intensity laser beams. The dominant emission direction ($\varphi_\ast,\vartheta_\ast$) of the signal photons $\gamma_{\ast}$ is highlighted by an arrow.}
 \label{fig:Setup}
\end{figure}

In scenario (i), the three beams collide perpendicularly, such that $\hat{\vec{e}}_{\kappa_1}=\hat{\vec{e}}_{\rm z}$, $\hat{\vec{e}}_{\kappa_2}=\hat{\vec{e}}_{\rm x}$ and $\hat{\vec{e}}_{\kappa_3}=\hat{\vec{e}}_{\rm y}$, where $\{\hat{\vec{e}}_{\rm x},\hat{\vec{e}}_{\rm y},\hat{\vec{e}}_{\rm z}\}$ span the spatial coordinate system. 
Following Ref.~\cite{Lundstrom:2005za}, we focus on the scenario where two frequency-doubled beams collide with a fundamental-frequency beam, and the polarization vectors of the beams are $\hat{\vec{e}}_{E_1}=\hat{\vec{e}}_{\rm y}$, $\hat{\vec{e}}_{E_2}=\hat{\vec{e}}_{\rm z}$ and $\hat{\vec{e}}_{E_3}=\hat{\vec{e}}_{\rm x}$.
The characteristics of the signal photons attainable in this set-up are summarized in Table~\ref{tab:Ntot2}; for an illustration of the collision geometry, see Figure~\ref{fig:Setup}.

We obtain $N\approx2.42$ signal photons per shot for the scenario based on the availability of two $1\,{\rm PW}$ lasers. For the design parameters of ELI-NP \cite{ELI} envisaging two $10\,{\rm PW}$ lasers, this number would even increase by a factor of $1000$. 
As the dominant scattering process is characterized by the absorption of two frequency-doubled laser photons and the emission of one low-energy fundamental-frequency laser photon, the signal photons exhibit a distinct frequency $\omega_\ast\approx3\omega$, different from the frequencies $\omega_1=\omega$ and $\omega_{\{2,3\}}=2\omega$ of the high-intensity lasers.
Moreover, they are emitted into a specific direction outside the forward cones of the laser beams driving the effect.
Both properties should allow for an excellent signal to background separation in experiment.

{For the present case, the main emission frequencies and directions can be inferred remarkably precisely from simplistic plane-wave-type considerations which characterize each beam by just two parameters: its frequency $\omega_b$ and its wave-vector $\vec{\kappa}_b=\omega_b\hat{\vec{e}}_{\kappa_b}$. Deviations from the plane-wave case occur, e.g., because of the finite pulse duration  $\tau_b$. }
{For the pulse duration of $\tau_b=25\,{\rm fs}$ adopted here and a beam of frequency $\omega_b=n\omega$, we expect corrections to be parametrically suppressed with $1/\tau_b\omega_b\approx 1/(58.5n)\ll 1$.
The Fourier transform of the beam's temporal profile in the focus reveals its energy spectrum  $\Omega$, $\int{\rm d}t\,{\rm e}^{{\rm i}\Omega t}{\rm e}^{-4(t/\tau_b)^2} \cos(\omega_b t)=\sqrt{\pi}\,(\tau_b/4)\sum_{s=\pm1}\exp\{-(\tau_b/4)^2(\Omega-s\omega_b)^2\}$; for $\tau_b\omega_b\gg1$, this spectrum is strongly peaked at the plane-wave frequencies $\Omega=\pm\omega_b$.
The plane-wave-model selection rules for the signal photon energy can hence be expected to hold to a very good accuracy for the present case; in turn, significant deviations may occur for shorter pulses.
On the other hand, for beams focused down to the diffraction limit as assumed here, the beam waist is as small as the laser wavelength, $w_{0,b}=\lambda_b$. This suggests that the strict three-momentum conservation involving photon wave-vectors only, as predicted by a plane-wave model, can be modified quite substantially by corrections depending on the additional scale $w_{0,b}$.
However, in combination with the photon on-shell condition, $|\vec{\kappa}_b|=\omega_b$, also these potential corrections may not deviate too severely from a plane-wave model. Under these conditions,
the predicted integrated numbers of signal photons should be compatible with plane-wave estimates.}

\setlength{\extrarowheight}{3pt}
\begin{table}[t]
 \caption{Exemplary results for a three dimensional collision geometry. The incident high-intensity laser beams $b$ (polarization vector $\hat{\vec{e}}_{E_b}$) propagate in direction $\hat{\vec{e}}_{\kappa_b}$.
 They are characterized by their pulse duration $\tau_b$, energy $W_b$ and frequency $\omega_b$ ($\omega=1.55 \, {\rm eV}$).
 We assume  $\tau_b=25\,{\rm fs}$ for all beams.
 The signal photons exhibit a distinct energy $\omega_{\ast}$ and are emitted in the direction $(\varphi_{\ast},\vartheta_{\ast})$ with a radial divergence of $\theta_\ast$.
 $N$ is the number of signal photons per shot.}
\begin{ruledtabular}
\begin{tabular}{lccc}
Beam parameters & $b=1$ & $b=2$ & $b=3$ \\
\hline
$W_b[{\rm J}]$ & 25 & 6.25 & 6.25 \\
$\omega_b$[$\omega$] & 1 & 2 & 2 \\
$\hat{\vec{e}}_{\kappa_b}$ & $\hat{\vec{e}}_{\rm z}$ & $\hat{\vec{e}}_{\rm x}$ & $\hat{\vec{e}}_{\rm y}$ \\
$\hat{\vec{e}}_{E_b}$ & $\hat{\vec{e}}_{\rm y}$ & $\hat{\vec{e}}_{\rm z}$ & $\hat{\vec{e}}_{\rm x}$ \\
\hline
\multicolumn{4}{l}{Signal photon characteristics} \\
$\omega_{\ast}$[$\omega$] & \multicolumn{3}{r}{3} \\
$\varphi_{\ast}$[$^{\circ}$] & \multicolumn{3}{r}{45} \\
$\vartheta_{\ast}$[$^{\circ}$] & \multicolumn{3}{r}{110.74} \\
$\theta_\ast$[$^{\circ}$] & \multicolumn{3}{r}{15.91} \\
$\boldsymbol{N}$ & \multicolumn{3}{r}{\textbf{2.42}} \\
\end{tabular}
\end{ruledtabular}
 \label{tab:Ntot2}
\end{table}
\setlength{\extrarowheight}{0pt}

{We quantify these considerations by comparing} the above result with the estimate provided by Ref.~\cite{Lundstrom:2005za},
based on a simplified modeling of the colliding laser beams as plane waves and indirectly accounting for finite beam width and focusing effects: Employing the design parameters of the Astra Gemini laser system \cite{AstraGemini}, and assuming all lasers to be focused down to the wavelength of the fundamental frequency laser, Ref.~\cite{Lundstrom:2005za} estimated the number of signal photons per shot as $0.07$.
Using the same parameters, but employing a realistic modeling of the high-intensity laser fields as pulsed Gaussian beams, we find a 70\% increase: $N\approx0.12$. We observe that the signal photon number can be further enhanced by a factor of almost 2 by focusing all three beams down to the diffraction limit with $f^{\#}=1$ yielding $N\approx0.23$.

{The reason for the quantitative difference between our first-principles description and the plane-wave model can be understood:}
Plane waves are infinitely extended, and thus not confined to any
prescribed space-time volume.  In turn, a modeling of the scattering
process of focused high-intensity laser pulses with
plane waves generically relies on several decisive {\it ad hoc} assumptions, e.g.,
that the plane waves are only interacting for a certain time interval, and all their photons are
contained within a given transverse extent.  
For instance, Ref.~\cite{Lundstrom:2005za} models the interaction
region as a cube.  The
field strength of the combined fields in the interaction region
is then obtained by assuming the lasers' energy to be concentrated in
this cube.  As a consequence, the lasers'
combined field strength is effectively smeared out over the interaction volume, resulting in an
effective field strength somewhat below the actual peak
field strength.  However, the number of attainable signal photons
depends decisively on the peak field strength of the superimposed
high-intensity laser fields in the interaction region; c.f., e.g.,
Ref.~\cite{wir}.  Higher
peak fields result in larger signal photon yields.  Plane-wave models
of focused high-intensity pulses inevitably involve such averaging
procedures, and thus tend to predict lower signal photon numbers.

\begin{figure}[ht]
\begin{center}
 \includegraphics[trim={0cm 0 0 0},clip,width=0.48\textwidth]{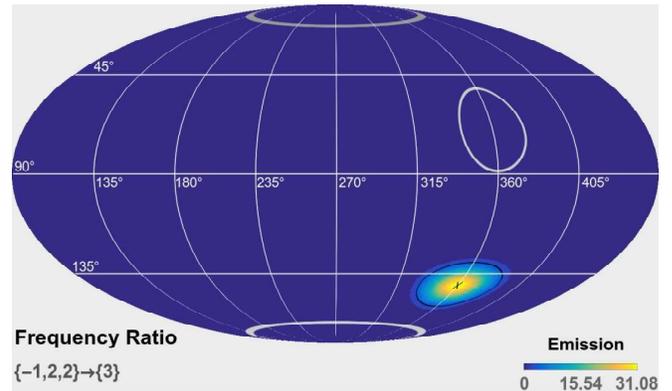}
\end{center} 
 \caption{Directional emission characteristics of signal photons for the scenario featured in the last line of Table~\ref{tab:Ntot} using the Mollweide projection. 
 The signal photons of energy $\omega_\ast \approx 3\omega = 4.65\,{\rm eV}$ are emitted predominantly in directions outside the forward cones of the high-intensity lasers, delimited by $\theta_b\approx18.24^\circ$ and
 depicted as light (dark) gray circles for the beams of frequency $2\omega$ ($\omega$).
 {The dominant emission direction lies in the collision plane and is marked by a cross. The signal falls off asymmetrically: Its radial divergence is well-approximated by an ellipsis (black) with minor and major radial divergences $\theta_{\ast}^\text{minor}\approx11.1^\circ$ and  $\theta_{\ast}^\text{major}\approx28.7^\circ$ in and perpendicular to the collision plane, respectively.}}
 \label{fig:Sph122b_M}
\end{figure}

{The improvement from an adequate spatiotemporal treatment of the laser pulse properties for accurate predictions becomes most obvious from the following fact:}
{the relaxation of the strict three-momentum conservation condition mentioned above gives rise to a finite divergence $\theta_\ast$ of the signal photons about the dominant emission direction which cannot be inferred from a plane-wave model.} {This observable can straightforwardly be computed from our method in a fully angle resolved manner. Quantitative results are given in Tab.~\ref{tab:Ntot2} and Fig.~\ref{fig:Sph122b_M}.}

In the second scenario, (ii), the beam axes of the lasers are confined to the $\rm x$-$\rm z$ plane.
All beams are polarized perpendicular to the collision plane, i.e., $\hat{\vec{e}}_{E_b}=\hat{\vec{e}}_{\rm y}$ for $b\in\{1,2,3\}$.
Without loss of generality we assume the first laser to propagate along the $\rm z$ axis, $\hat{\vec{e}}_{\kappa_1}=\hat{\vec{e}}_{\rm z}$, and parameterize the beam axes of the other lasers as $\hat{\vec{e}}_{\kappa_b}=\sin\vartheta_b\,\hat{\vec{e}}_{\rm x}+\cos\vartheta_b\,\hat{\vec{e}}_{\rm z}$. The angle $\vartheta_b$ measures the inclination of beam $b\in\{2,3\}$ relative to the first.

Configurations of this type have originally been studied in Refs.~\cite{Lundstrom:2005za,Gies:2016czm}, based on a simplified description treating all \cite{Lundstrom:2005za}, or all but one \cite{Gies:2016czm} high-intensity laser beams as plane waves.
In the present letter we go a significant step beyond: Based on the
well-controlled approximations given above, we provide quantitative predictions of the
numbers of signal photons attainable in a realistic experiment from
first principles for the first time.

\setlength{\extrarowheight}{3pt}
\begin{table}[t]
 \caption{Signal photon yield $N$ for various collision geometries.
 The beam axes of the three lasers (energy $W_b$, frequency $\omega_b$, pulse duration $\tau_1=\tau_2=\tau_3=25\,{\rm fs}$) driving the effect are confined to the $\rm x$-$\rm z$ plane, and are polarized along $\rm y$, i.e., $\hat{\vec{e}}_{E_b}=\hat{\vec{e}}_{\rm y}$ for all beams.
 The laser frequencies $\omega_b$ are given in multiples of $\omega=1.55\,{\rm eV}$.
 The signs indicate the dominant signal photon emission channel, where $+(-)$ stands for absorbed (released) laser photons in the microscopic interaction process.
 Correspondingly, the dominant signal photon emission frequency can be inferred as $\omega_\ast=\sum_{b=1}^3 \omega_b$.
 The angle $\vartheta_2$ ($\vartheta_3$) measures the inclination of the 2nd (3rd) beam with respect to the first.}
\begin{ruledtabular}
\begin{tabular}{ccccccccc}
$W_1[{\rm J}]$ & $W_2[{\rm J}]$ & $W_3[{\rm J}]$ & $\omega_1[\omega]$  & $\omega_2[\omega]$ & $\omega_3[\omega]$ & $\vartheta_2$[$^{\circ}$] & $\vartheta_3$[$^{\circ}$] & $\boldsymbol{N}$ \\
\hline
25.0 & 12.5 & 12.5 & 1 & -1 & 1 & 90 & 180 & \textbf{4.90} \\
12.5 & 12.5 & 12.5 & 2 & -1 & 1 & 90 & 216 & \textbf{4.03} \\
12.5 & 12.5 & 6.25 & 2 & -1 & 2 & 119.75 & 239.5 & \textbf{3.99} \\
25.0 & 6.25 & 6.25 & -1 & 2 & 2 & 70.47 & 180 & \textbf{3.03} 
\end{tabular}
\end{ruledtabular}
 \label{tab:Ntot}
\end{table}
\setlength{\extrarowheight}{0pt} 

Table~\ref{tab:Ntot} summarizes the maximum attainable numbers of signal photons $N$ per shot for various choices of the beam frequencies $\omega_b$, together with the corresponding pulse energies $W_b$, keeping the pulse durations $\tau_b=25\,{\rm fs}$ fixed for all beams.
To arrive at these results, for given laser parameters the angles $\vartheta_2$ and $\vartheta_3$ are adjusted such that the value of $N$ becomes maximal.
Note, that both scenarios (i) and (ii) yield comparable numbers of signal photons. Particularly, for the same laser parameters (cf. Tab.~\ref{tab:Ntot2} and last line of Tab.~\ref{tab:Ntot}), we obtain $N\approx2.42$ and $N\approx3.03$ for scenario (i) and (ii), respectively.
We emphasize once again that an upscaling of the available laser intensity to $10\,{\rm PW}$ class lasers would increase the signal photon yield by a factor of $1000$.

A measurement of this fundamental phenomenon will clearly be challenging, because of the large photon background of the incoming beams. Still, our scenario offers a substantial set of lever arms that facilitate an unambiguous identification of the signal photons: (a) the source of the signal photons is precisely localized in spacetime, (b) the directional signal photon emission characteristics can be arranged to lie far outside the forward cones of the high-intensity beams delimited by $\theta_n$, see Fig.~\ref{fig:Sph122b_M}, (c) the distinct signal frequency of $\omega_\ast\approx3\omega$ outside the frequencies of the driving laser beams $\omega_b\in\{\omega,2\omega\}$ facilitates strong background suppression by frequency filtering, (d) the background can be studied in detail in advance for each beam as well as for mutual two-beam collisions, (e) for suitable polarization configurations, the signal polarization can allow for further polarization filtering, {(f) the predicted signal beam divergence can help optimizing the detector design}. 

\paragraph{Conclusions}

In this letter, we have studied the collision of three realistically modeled laser pulses in vacuo in unprecedented detail. 
More specifically, we have modeled the high-intensity laser fields as experimentally realistic pulsed Gaussian beams, thereby significantly advancing beyond previous studies. 
Our results substantiate previous estimates suggesting the possibility of measuring signatures of QED vacuum nonlinearity with state-of-the-art technology.
The predictive power of our method arises from reformulating the effective, fluctuation-mediated interaction process as a vacuum emission process, giving rise to signal photons with characteristic kinematic properties encoding the signature of quantum vacuum nonlinearity.
Upon combination with an efficient numerical algorithm, this approach facilitates quantitatively accurate predictions of the numbers of attainable signal photons and their kinematic characteristics in experiment. Our results suggest that a first discovery experiment of nonlinear interactions of macroscopically controllable electromagnetic fields is in reach with the present generation of high-intensity lasers coming online just now in many laser labs worldwide -- decades after the seminal work of Refs.~\cite{Euler:1935zz,Heisenberg:1935qt,Weisskopf}.

\begin{acknowledgments}
  We are grateful to Andr\'{e} Sternbeck and Matt Zepf for helpful discussions. The
  work of C.K. is funded by the Helmholtz Association through the
  Helmholtz Postdoc Programme (PD-316).  We acknowledge support by the
  BMBF under grant No. 05P15SJFAA (FAIR-APPA-SPARC).  Computations
  were performed on the ``Supermicro Server 1028TR-TF'' in Jena, which
  was funded by the Helmholtz Postdoc Programme (PD-316).
\end{acknowledgments}

\end{document}